# Performance Evaluation of Cryptographic Ciphers on IoT Devices

Kedar Deshpande
Department of Telecommunication
Sir M Visvesvaraya Institute of Technology
Bangalore, India
vvkedarvv@gmail.com

Praneet Singh
Department of Electronics and Communication
Ramaiah Institute of Technology
Bangalore, India
praneet195@gmail.com

*Abstract*—With the advent of Internet of Things (IoT) and the increasing use of application-based processors, security infrastructure needs to be examined on some widely-used IoT hardware architectures. Applications in today's world are moving towards IoT concepts as this makes them fast, efficient, modular and future-proof. However, this leads to a greater security risk as IoT devices thrive in an ecosystem of co-existence and interconnection. As a result of these security risks, it is of utmost importance to test the existing cryptographic ciphers on such devices and determine if they are viable in terms of swiftness of execution time and memory consumption efficiency. It is also important to determine if there is a requirement to develop new lightweight cryptographic ciphers for these devices. This paper hopes to accomplish the above- mentioned objective by testing various encryption-decryption techniques on different IoT based devices and creating a comparison of execution speeds between these devices for a variety of different data sizes.

## I. INTRODUCTION

The science of securing digital data by making it unintelligible for unauthorized access, especially for transmission and storage is called Cryptography. The usage of Internet of Things and other applications has led to an exponential increase in the data being stored, transmitted and processed. This data increase has led to an increased demand for data security architecture. Many applications of IoT are based out of application processors, the common ones being Raspberry Pi and Beagle Bone. References [1] and [2] show us the different IoT based applications of these devices. However, employing security mechanisms on such processors will lead to an overload in the already loaded processors. This may result in increased power consumption, application delays or increased resource demands. As a result, there is a necessity to examine the various symmetric and asymmetric encryption-decryption techniques available and test their effects on the Raspberry Pi 3 and Beagle Bone Black processors and compare various parameters like speed and efficiency. This would also help us determine the need of light-weight schemes on such devie The various security techniques we are going to compare in this paper are: Twofish, Blowfish, DES, Triple-DES, AES, RC2, RC4 and ChaCha20. These ciphers are tested on the IoT devices by running them on different file sizes ranging from 1 MB to 128 MB.

## II. IoT DEVICES

### A. Raspberry Pi 3

The Raspberry Pi family consists of pocket-sized computers containing high memory, fast processor and various ports which can be interfaced with a large ecosystem of devices as required by the application. Reference [3] gives us an excellent comparison between a member of this family and various other IoT devices. These devices are also backed by a large community for support.

Raspberry Pi 3 is the latest of the editions on which we will run our security mechanisms. It has dimensions of 8.7cm in length, 5.8cm in width and 1.8cm in height. It is powered by a Quad-core Broadcom BCM2837 64-bit CPU clocked at 1.2 Ghz. Its 1GB DDR2 RAM makes it suitable and speedy for IoT based applications. It also comes with a built-in BCM43438 wireless LAN and low energy Bluetooth chip on board. It contains a wide array of ports such as 26 GPIO pins, 4 USB2.0 ports, 4 Pole audio stereo output and a full-sized HDMI port. It also contains a CSI camera port which can be used quickly and efficiently for video analytics-based applications. Raspbian Sketch is the preferred operating system. A 3A charger at 5V is sufficient to power this device.

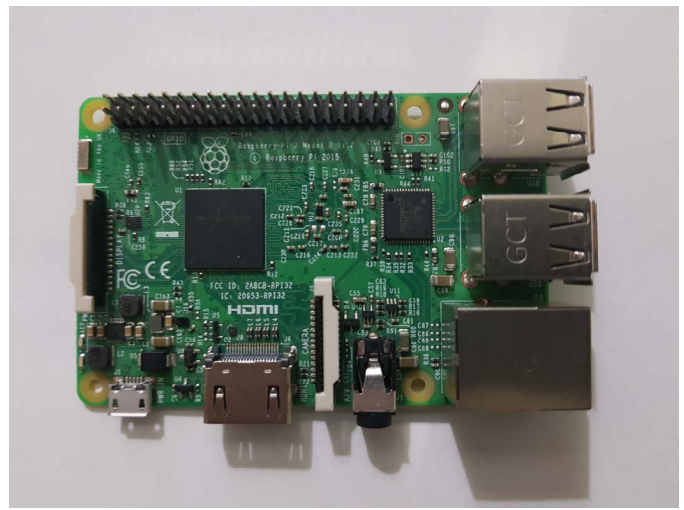

Figure 1. Board Representation of Raspberry Pi 3

TABLE I.

Comparison Between Raspberry Pi 3 and Beagle Bone Black

| Parameters | Raspberry Pi 3 | Beagle Bone Black |
|---|---|---|
| SoC | Broadcom BCM2837 | TI Sitara AM335x |
| Processor | 1.2Ghz 64-bit ARM Cortex-A53 | 1Ghz 32-bit ARM Cortex-A8 |
| RAM | 1GB DDR2 | 512MB DDR3 |
| GPU | VideoCore IV | PowerVR SGX530 |
| Operating System | Raspbian Stretch 4.9 | Debian Wheezy 9.0 |
| GPIO Pins | 26 | 65 |
| Power Consumption | 210-460 mA @ 5V under varying conditions | 150-350 mA @ 5V under varying conditions |
| Storage | On-Board 4GB storage extendable via Micro SD card | Limited to Micro SD card storage |
| Features | On-Board Bluetooth and Wi-Fi cards, HDMI port, USB Ports, Ethernet Port, 3.5mm Audio Jack | USB Port, Micro HDMI Port, Ethernet Port |

### B. Beagle Bone

Homogeneous to the Raspberry Pi, the Beagle Bone family consists of multi-purpose hardware with an array of features and ports. Providing additional GPIO functionality over the Raspberry Pi 3, these devices are widespread in the field of IoT. Reference [4] provides a detailed description of various devices belonging to this family.

With dimensions of 8.62cm in length, 5.33cm in width and 1.6cm in height, the Beagle Bone Black is slightly power deficient when compared to the Raspberry Pi 3. It is packed with a 4GB 8-bit embedded multimedia controller onboard flash storage and a 512MB DDR3 RAM. It is provided with 3D graphics and NEOB floating-point accelerator and 2, 32-bit programmable real-time units each clocked at 200 Mhz. It has a USB host, Ethernet port, a micro HDMI port and 2 46-pin headers. It runs on the Debian Wheezy 9. The main advantage for the Beagle Bone Black over the Raspberry Pi 3 is the availability of 65 GPIO pins, thus providing the user with better functionality and control over run-time. References [5] and [6] shows applications of this device as well as helps explain the device's architecture in more detail.

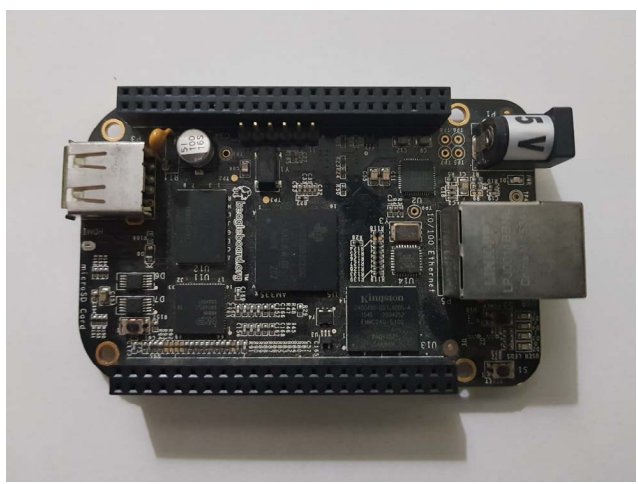

Figure 2. Board Representation of Beagle Bone Black

Table I compares the two IoT devices that are being tested in this paper and it states similar differences as seen in [3].

### III. CRYPTOGRAPHIC CIPHERS

With the development of security schemes over the years, many new encryption techniques have been devised, and improvements have been done on existing techniques. In general, all the existing techniques can be classified into Asymmetric and Symmetric encryption techniques. In this paper, we will be concentrating mainly on the widely used Symmetric encryption techniques. A detailed analysis, working and the various attacks on these Symmetric and Asymmetric ciphers are seen in [7].

Symmetric encryption techniques are further classified into Block Ciphers and Stream Ciphers. The block and stream ciphers that have been used in this paper are discussed next.

#### A. Stream Ciphers

Stream Cipher algorithms peruse the entire intelligible message and convert each symbol of the plain text directly into a symbol of cipher text. The symbol is generally a bit, and the transformation performed is generally exclusive-OR (XOR). Due to bit by bit encoding, they are lighter and faster schemes relying solely on confusion concepts. They also have statistically random structures and are easier to implement on hardware. Reference [8] discusses a few attacks on stream ciphers. We will talk about the 2 most prominent, fast and light stream ciphers, RC-4 and ChaCha 20, and compare them on the Beagle Bone and Raspberry Pi.

- Rivest Cipher 4 (RC4)

Rivest Cipher 4 abbreviated as RC4 was developed by Ronald Rivest in 1987. It relies on a symmetric key algorithm to generate a keystream sequence for encryption and decryption. The data stream is simply XOR-ed with the generated key sequence. A detailed analysis of Rivest Ciphers is performed in [9].

The key generation algorithm is completely individualistic of plain text, and the key length variable, with the maximum length being 256 bytes. The algorithm uses a 256 byte array called S. This S array is initialized to permutations of 0 to 255 using a Key-Scheduling algorithm. These values in the S array are then processed for 256 iterations to form a random combination of the permutation values. The Pseudo-Random generation algorithm is then used to further modify the output of each key byte and XOR key bytes with plain text bytes or vice versa. A lookup stage of RC4 is as shown in Fig. 3.

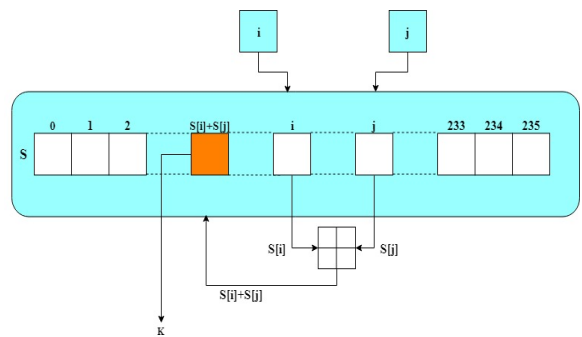

Figure 3. Lookup stage of RC4 stream cipher

- ChaCha20

    Developed by Daniel J. Bernstein, the ChaCha 20 stream cipher is a variant of the Salsa20 stream cipher. The design principles of this cipher are almost identical to that of Salsa20, however, there is increased diffusion per round. Reference [10] gives us an exhaustive explanation of the algorithm by D.J. Bernstein himself.

    It is a 256 bit stream cipher. The changes from Salsa20/8 to ChaCha8 are designed to improve diffusion per round, thus increasing resistance to cryptanalysis, while preserving and improving time per round. However, the extra diffusion does not add more operations when compared to Salsa20. A ChaCha round has 16 additions and 16 xors and 16 constant-distance rotations of 32-bit words. The parallelism and vectorizability of the ChaCha 20 algorithm are conformant with that of Salsa20.

B. *Block Ciphers*

    Block Cipher cryptographic schemes convert an entire block of plain text into a block of cipher text at a time. These are bulkier and slower ciphers as they involve the division of plain text into blocks and rely on both diffusion and confusion concepts. They have a simpler software implementation and also have different modes of operations. The block ciphers discussed and used in this paper have been run on the two simplest and fastest modes which are Cipher Block Chaining (CBC) and Electronic Code Book Mode (ECB). The various other operation modes can be seen in [11]. A detailed analysis of the various attacks on these ciphers is seen in [12]. The block ciphers discussed in this paper are based on the Feistel cipher structure as showin in Fig. 4. The different block ciphers discussed in this paper are AES, DES, Triple-DES, RC2, Blowfish and Twofish ciphers. The block ciphers that have been used in this paper are discussed next.

- Advanced Encryption Standard (AES)

    Advanced Encryption Standard or AES is a block encryption technique which was developed by Belgian cryptographers, Vincent Rijmen and Joan Daemen. It is based on the principle of substitution-permutation network, a combination of both substitution and combination. It basically comprises of 3 block ciphers- AES-128, AES-192 AES-256 and each of these ciphers can encrypt and decrypt data in 128-bit blocks using 128, 192 and 256 bit keys respectively. The higher the key size, the stronger the encryption. Since AES is a symmetric cipher, both the sender and the receiver must know the key for encryption and decryption respectively.

AES defines 4 transformations to convert the plain text into cipher text. The first step involves arranging data into an array or matrix. The second step shifts data rows, the third step mixes columns and the last step performs simple XOR operation on each column using a different part of the encryption key. 10 such rounds are performed for 128-bit keys, 12 rounds for 192-bit keys and 14 rounds for 256-bit keys. Reference [13] provides a detailed insight of this cipher.

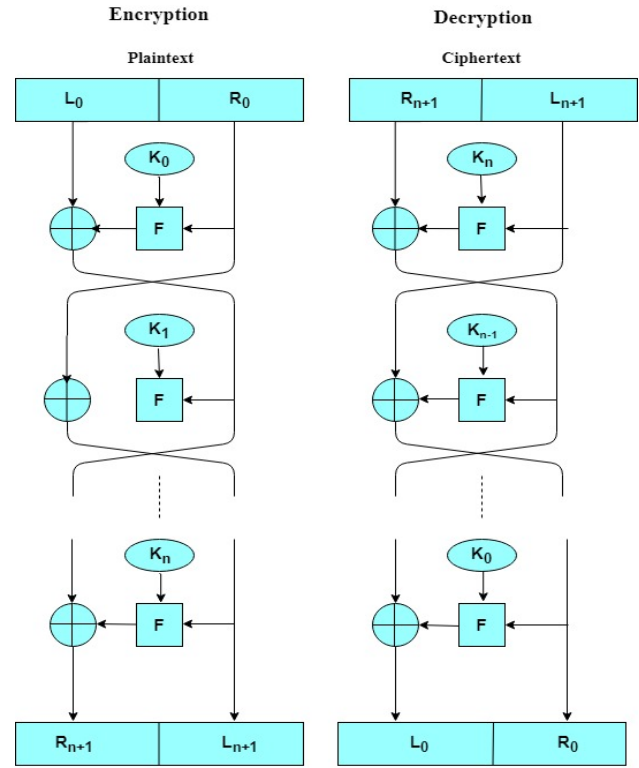

Figure 4. Feistel Cipher Structure

- Data Encryption Standard (DES)

    Data Encryption Standard or DES was developed in 1970 by IBM. It is a block cipher that takes in 64-bit plaintext and after a series of operations, converts it into a 64-bit cipher text. DES is a symmetric cipher and uses a key for these operations of length 64-bits, out of which 56 bits are used for encryption-decryption and the remaining 8 bits are used to check parity. Thus, DES has an effective key length of 56 bits. The algorithm consists of 16 identical rounds. A thorough analysis and working of this cipher is seen in [13] as well.

    Initially, the 64-bit plain text is divided into two 32-bit blocks. These 2 blocks are processed separately in each of the 16 rounds. This structure is referred to as the Feistel Structure. The F-block in the structure scrambles a half block with some part of the key, whose output is combined with the other half block. These 2 halves are swapped before the next round. The Initial Permutation (IP) and Final permutation rounds are inverses of each other. Being a symmetric cipher, it uses the same key for encryption is used for decryption, but in the reverse order. This makes it easier to design hardware and software for encryption and decryption. A detailed comparison between AES and DES is also seen in [13].

- Triple-Data Encryption Standard (3DES)

    Triple Data Encryption Standard or 3DES algorithm basically runs the DES algorithm 3 times on a given plaintext. The original DES's 56-bit key was sufficient to provide security but the availability of additional

computational power led to increased brute-force attacks. This led to the development of the 3DES cipher.

3DES uses a 168 bit key and operates on a block size of 64-bits. Although more secure than the former DES algorithm, it is found to be the one of the mostslowest block cipher in existence due to its excessive computational complexity. An all-inclusive explanation and detailed analysis of this cipher is seen in [14].

- Blowfish

Blowfish block cipher was developed 1993 by Bruce Schneier. It uses a fixed block of size 64 bits, with a varying key-length between 32 and 448 bits. It also makes use of large key-dependent S-boxes. Similar to DES, it has a 16-round Feistel cipher structure. It is an open source algorithm which has not yet been broken. It is also one of the fastest ciphers in public use. Reference [15] gives us an all-inclusive analysis and security enhancement for this cipher.

- Twofish

Similar to AES, DES and Blowfish algorithms, Twofish also depends on the Feistel structure. Having developed Blowfish, Bruce Schneier made developments to his cipher which thus lead to Twofish which is a symmetric cipher, with a block size of 128 bits and a key of any length upto 256 bits. The plain text is broken into two 32-bit words and fed into the F-boxes. Thw two words are further broken down into four bytes within these F-boxes and sent through S-boxes, each dependent on different keys. The four output bytes are combined into a 32-bit word using Maximum Distance Separable (MDS) matrix. The Pseudo-Hadamard Transform (PHT) is used to combine the 2 32-bit words. This is then XOR-ed with the other half. Certain 1-bit rotation operations are also performed before and after the XOR operation. The superiority of this cipher over the Blowfish cipher is seen in [16].

- Rivest Cipher 2 (RC2)

Taking inspiration from the RC4, Ronald Rivest in 1987 developed the Rivest Cipher 2. Abbreviated as RC2, it is a symmetric 64-bit block cipher with a variable key length of up to 128 bits. A brief explanation states that it involves a complicated round of operations to convert the plain text into cipher text. Based on a variable-length input key, a key-expansion algorithm is used to convert it into a fixed 64-bit key. This is followed by a sequence of operations involving 5 mixing rounds, a mashing round, 6 mixing rounds, another mashing round followed by another 5 mashing rounds.

A mixing round consists of 4 mix-up transformations. A round is said to be mashed by adding it to any one of the 16-bit words of the expanded key. A thorough comparsion of RC2 with other Rivest block ciphers is seen in [9].

IV. EXPERIMENTAL RESULTS

References [17] and [18] give us a detailed evaluation of the performance, efficiency and swiftness of block and stream cryptographic ciphers on commonly used Intel processors. However, these evaluations would not stand true for the IoT domain and as a result a similar evaluation is performed here. The cryptographic block and stream ciphers discussed in this paper were run on the Beagle Bone Black and Raspberry PI 3 for different data file sizes ranging from 1 MB to 128 MB to determine execution speed and time.

The key and block sizes for the various block ciphers are as shown in Table II.

TABLE II.
Key Sizes and Block Sizes for Block & Stream Ciphers

| Cipher | Key Size(bits) | Block Size(bits) |
|---|---|---|
| AES | 256 | 128 |
| DES | 56 | 64 |
| 3-DES | 168 | 64 |
| Blowfish | 128 | 64 |
| Twofish | 256 | 128 |
| RC2 | 128 | 64 |
| RC4 | 256 | - |
| ChaCha20 | 256 | - |

The execution time in second for various stream ciphers and block ciphers on the Rapsberry Pi 3 are as shown in Table III,Table IV and Table V .

TABLE III.
Block Cipher Executions in ECB Mode on Raspberry Pi 3

| File Size(MB) | Execution Time for Block Ciphers (s) | | | | | |
|---|---|---|---|---|---|---|
| | AES | DES | Triple-DES | Blowfish | Twofish | RC2 |
| 1 | 0.312512 | 0.340801 | 0.507857 | 0.303464 | 0.275164 | 0.354988 |
| 2 | 0.625167 | 0.680893 | 1.015655 | 0.603668 | 0.549241 | 0.713949 |
| 4 | 1.289868 | 1.405361 | 2.076549 | 1.251772 | 1.141539 | 1.467976 |
| 8 | 2.825549 | 3.058405 | 4.410864 | 2.749132 | 2.527144 | 3.180131 |
| 16 | 6.563713 | 7.078505 | 9.764627 | 6.432269 | 5.979014 | 7.707962 |
| 32 | 17.05146 | 19.95651 | 24.90283 | 16.69822 | 15.83249 | 18.40725 |
| 64 | 34.61057 | 36.42658 | 47.16157 | 33.66723 | 32.66352 | 37.00312 |
| 128 | 67.51322 | 72.89622 | 94.89921 | 66.73726 | 63.80815 | 73.98863 |

TABLE IV.
Block Cipher Executions in CBC Mode on Raspberry Pi 3

| File Size(MB) | Execution Time for Block Ciphers (s) | | | | | |
|---|---|---|---|---|---|---|
| | AES | DES | Triple-DES | Blowfish | Twofish | RC2 |
| 1 | 0.309982 | 0.332839 | 0.495875 | 0.291938 | 0.274767 | 0.354896 |
| 2 | 0.614113 | 0.663321 | 0.987876 | 0.579845 | 0.547751 | 0.707328 |
| 4 | 1.280448 | 1.372637 | 2.023355 | 1.207192 | 1.144121 | 1.466432 |
| 8 | 2.812099 | 2.992806 | 4.305038 | 2.660793 | 2.547753 | 3.164056 |
| 16 | 6.538891 | 6.915692 | 9.526417 | 6.620954 | 6.014227 | 7.265833 |
| 32 | 16.96479 | 17.67588 | 22.89434 | 17.10438 | 15.89528 | 18.34590 |
| 64 | 33.87311 | 36.42899 | 44.99781 | 33.00132 | 31.89973 | 36.98867 |
| 128 | 64.01567 | 72.66394 | 90.72483 | 63.98877 | 61.11897 | 72.89773 |

TABLE VI.
Stream Cipher Executions on Raspberry Pi 3

| File Size(MB) | Execution Time for Stream Ciphers (s) | |
|---|---|---|
| | RC4 | ChaCha20 |
| 1 | 0.241359 | 0.235922 |
| 2 | 0.500513 | 0.472921 |
| 4 | 1.040137 | 0.986634 |
| 8 | 2.328652 | 2.22117 |
| 16 | 5.585328 | 5.358924 |
| 32 | 14.998415 | 14.559962 |
| 64 | 30.553217 | 29.343859 |
| 128 | 61.883621 | 57.66328 |

The values for execution of the various stream ciphers and block ciphers on the Beagle Bone Black are as shown in Table VI, Table VII and Table VIII.

TABLE VI.
Block Cipher Executions in ECB Mode on Beagle Bone Black

| File Size(MB) | Execution Time for Block Ciphers (s) | | | | | |
|---|---|---|---|---|---|---|
| | AES | DES | Triple-DES | Blowfish | Twofish | RC2 |
| 1 | 0.005621 | 0.005811 | 0.006341 | 0.006175 | 0.005686 | 0.006138 |
| 2 | 1.166823 | 1.226297 | 1.568032 | 1.157565 | 1.103905 | 1.291024 |
| 4 | 2.565335 | 2.598819 | 3.273489 | 2.433813 | 2.362531 | 2.737681 |
| 8 | 5.954885 | 5.976521 | 7.391246 | 5.760432 | 5.612367 | 6.462823 |
| 16 | 14.89601 | 15.20358 | 17.87470 | 14.77387 | 14.56606 | 16.08261 |
| 32 | 30.61435 | 32.78581 | 37.18605 | 30.26231 | 30.18445 | 33.67839 |
| 64 | 57.63497 | 62.97124 | 72.67129 | 59.66723 | 57.66352 | 66.00312 |
| 128 | 110.4381 | 121.5390 | 138.7345 | 113.4342 | 110.3987 | 132.7962 |

TABLE VII.
Block Cipher Executions in ECB Mode on Beagle Bone Black

| File Size(MB) | Execution Time for Block Ciphers (s) | | | | | |
|---|---|---|---|---|---|---|
| | AES | DES | Triple-DES | Blowfish | Twofish | RC2 |
| 1 | 0.005831 | 0.005867 | 0.005941 | 0.005863 | 0.005732 | 0.005858 |
| 2 | 1.215087 | 1.250543 | 1.611351 | 1.559371 | 1.126454 | 1.324491 |
| 4 | 2.588259 | 2.693054 | 3.366573 | 2.463152 | 2.412783 | 2.79931 |
| 8 | 6.066367 | 6.205534 | 7.636621 | 5.845305 | 5.730861 | 6.499352 |
| 16 | 15.49762 | 15.67190 | 18.61863 | 14.99747 | 14.76463 | 16.26645 |
| 32 | 30.99188 | 32.99681 | 38.81971 | 30.41812 | 30.49481 | 33.17583 |
| 64 | 58.09342 | 63.98743 | 72.99781 | 59.43822 | 57.98245 | 68.79832 |
| 128 | 113.7991 | 123.5723 | 138.9981 | 117.4678 | 114.9821 | 133.8843 |

TABLE VIII.
Stream Cipher Executions on Beagle Bone Black

| File Size(MB) | Execution Time for Stream Ciphers (s) | |
|---|---|---|
| | RC4 | ChaCha20 |
| 1 | 0.00506499 | 0.00589123 |
| 2 | 1.075367 | 1.041657 |
| 4 | 2.302807 | 2.230733 |
| 8 | 5.498168 | 5.353476 |
| 16 | 14.355075 | 14.068853 |
| 32 | 29.275166 | 28.645975 |
| 64 | 57.363625 | 55.73215 |
| 128 | 112.001322 | 109.138972 |

Fig. 5, 6 and 7 show graphs comparing the speeds of the various block ciphers and stream ciphers on the Raspberry Pi 3 and Beagle Bone Black. We can see the variation of speeds for different file sizes in these graphs for the two devices being used.

It can be clearly inferred from the tabulated values for the Raspberry Pi 3 and the Beagle Bone Black that the Twofish algorithm has the highest speed amongst all the block ciphers. However both the stream ciphers, being light and fast compete with the Twofish algorithm. The ChaCha 20 stream cipher is clearly the most light, fast and efficient cipher amongst the ones discussed that can be run on the IoT devices.

Also it was seen that the CPU and memory consumption on the Beagle Bone Black averaged about 70 percent for the various encryption schemes. However the Raspberry Pi executed all the schemes with an average memory consumption of 40 percent which is much lower then the Beagle Bone Black.

However, as seen in [19] and [20], several light weight ciphers have been developed which compete with the fastest cipher seen here in terms of speed and also use fewer memory resources on such devices.

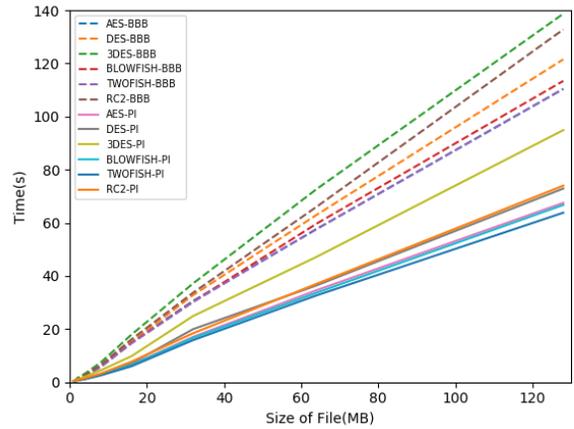

Figure 5. Execution Speed Comparison of Block Ciphers in ECB Mode between Raspberry Pi 3 and Beagle Bone Black

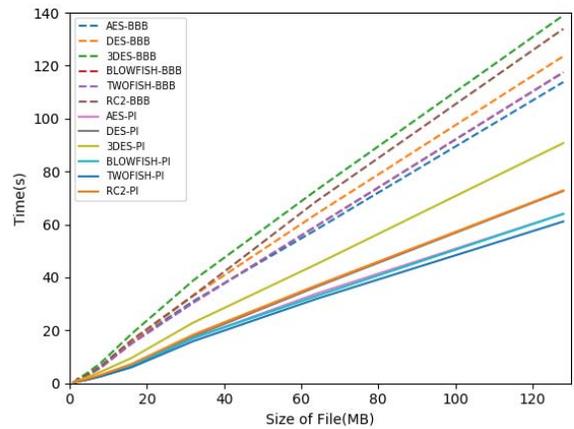

Figure 6. Execution Speed Comparison of Block Ciphers in CBC Mode between Raspberry Pi 3 and Beagle Bone Black

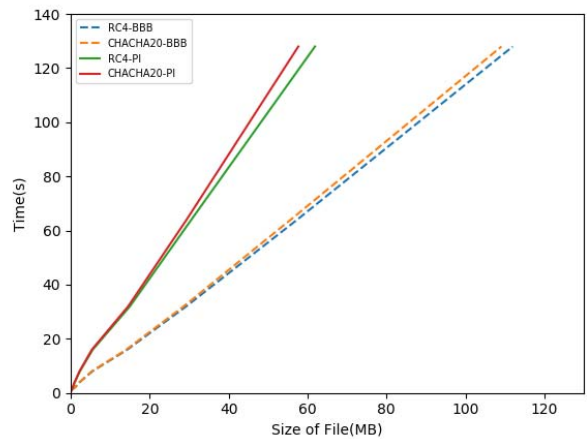

Figure 7. Execution Speed Comparison of Stream Ciphers between Raspberry Pi 3 and Beagle Bone Black

## IV. Conclusion

We have tested the two most competitive IoT devices and compared there performance results. Due to the processing speed on the Beagle Bone Black being lower than that of the Raspberry Pi 3, the execution time of these ciphers nearly doubles on it. The power and memory consumption was also found to be lower on the Raspberry Pi 3. As a result, for quick, efficient, secure and fast data transmission the Raspberry Pi 3 performs better than the Beagle Bone Black. However, if several interfaces need to be added on as seen in several IoT applications, the Beagle Bone Black has better available functionality with its replete GPIO pins.

The next step in the development of cryptographic ciphers for IoT is to either refine the existing ciphers or develop new light weight schemes which would help in improving the performance and memory consumption for these IoT devices.